\documentclass[prl,twocolumn]{revtex4-1}

\usepackage{graphicx}
\usepackage{multirow}
\usepackage{dcolumn}
\usepackage{bm}
\usepackage[normalem]{ulem}
\usepackage{amsmath}
\usepackage{amsfonts}
\usepackage{amssymb}
\usepackage{color}
\usepackage{colordvi}
\usepackage{dcolumn}
\newcommand{\beq}{\begin{equation}}
\newcommand{\eeq}{\end{equation}}
\newcommand{\bea}{\begin{eqnarray}}
\newcommand{\eea}{\end{eqnarray}}

\usepackage[colorlinks=true, breaklinks=true, linkcolor=blue,
urlcolor=blue, citecolor=blue]{hyperref}

\begin{document}

\title{Effects of 
 Lifshitz 
  transitions in ferromagnetic superconductors: the case of URhGe}
\author {Yury Sherkunov$^1$, Andrey V. Chubukov$^2$ and Joseph J. Betouras$^1$}
\affiliation {$^1$Department of Physics and Centre for the Science of Materials, Loughborough University, Loughborough, LE11 3TU, United Kingdom}
\affiliation {$^2$Department of Physics, University of Minnesota, Minneapolis, Minnesota 55455, USA}
\date{\today}

\begin{abstract}

In ferromagnetic superconductors, like URhGe,  superconductivity co-exists with magnetism  near zero field, but then re-appears again in a finite field range, where the system also displays mass enhancement in the normal state.
We present theoretical understanding of
 this non-monotonic behavior.
 We explore the multi-band nature of URhGe and associate re-entrant superconductivity and mass enhancement with the finite field
    Lifshitz transition in one of the bands.  We found
     good
     agreement between our theory
    and a number of experimental results for URhGe,  such as weakly first order reentrant transition,
    the dependence of superconducting T$_c$ on a
      magnetic field, and the field dependence of the effective mass, the specific heat and the resistivity in the normal state.  Our theory can be
      applied to other ferromagnetic multi-band superconductors.
\end{abstract}

\maketitle

Ferromagnetic superconductors are exciting systems to study the interplay of magnetism and superconductivity, contrary to the common wisdom that
the presence of a ferromagnetic order destroys superconductivity. The coexistence of superconductivity and ferromagnetism has been realised experimentally for uranium-based heavy-fermion compounds,
  like UGe$_2$ \cite{Saxena00}, UCoGe \cite{Huy07} and URhGe \cite{Aoki01}. The materials exhibit a wealth of exotic properties, including, e.g.,
  the appearance of non-Landau damping in magnetic excitations \cite{ChubukovBetourasEfremov}.

\begin{figure}[h]
\includegraphics[width=0.45\textwidth]{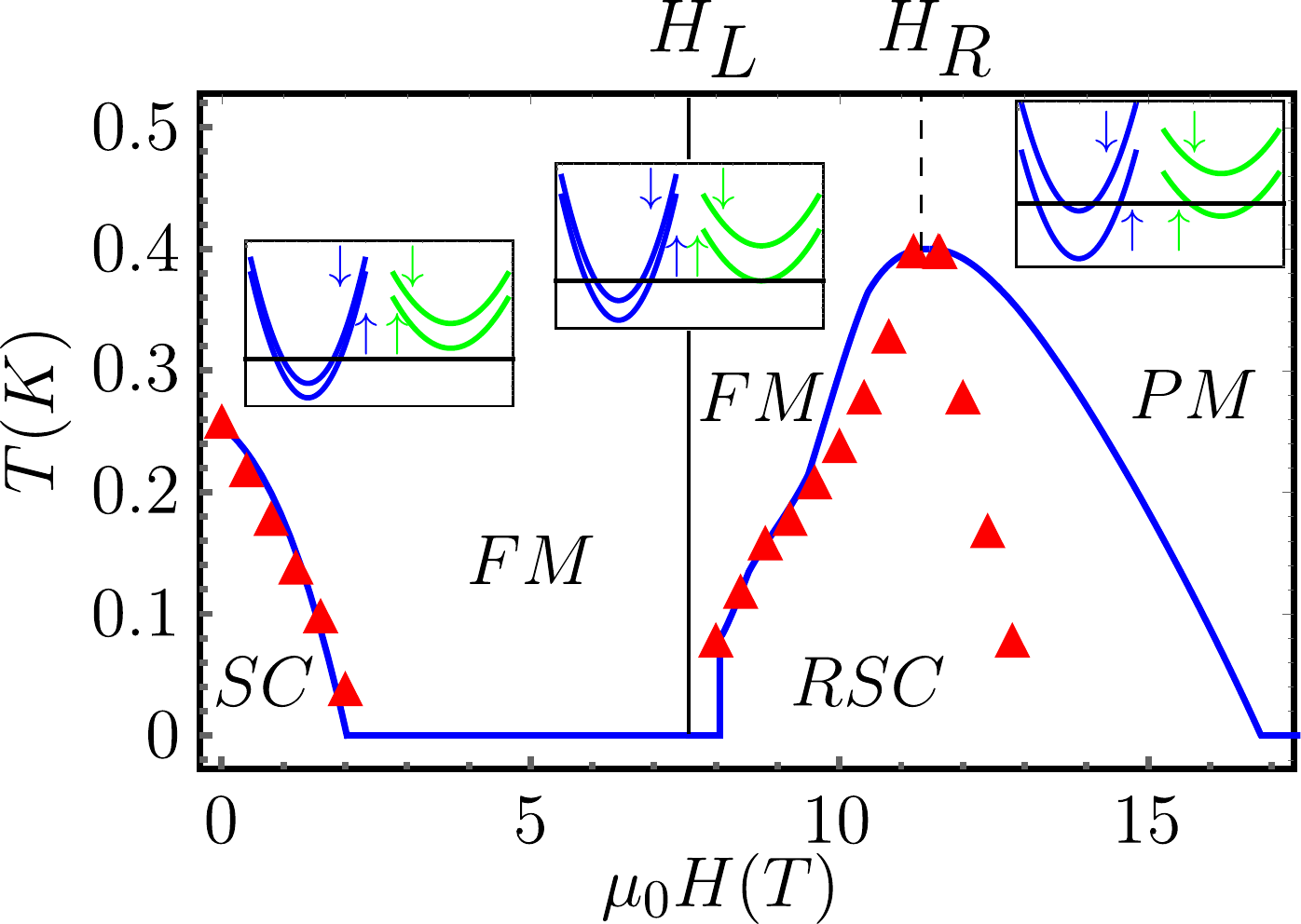}
\caption{The experimental and theoretical diagrams of URhGe in an external magnetic field $H$.
   Triangles are the experimental data
      ~\cite{LevyScience05},
     solid
        line is the theoretical  result.
          Superconductivity is present near $H=0$, absent at intermediate fields, and re-appears at higher fields, with a maximum at around $10-11$T. In the insets, we show the fermionic dispersion in our model of two electron bands, separated by $\mathbf K_0$ in the momentum space and exhibiting Zeeman splitting. The
            Lifshitz transition occurs at $H_L$, when
            the spin-up branch of band 2
               touches the chemical potential.  At higher $H$, this band opens up a new Fermi surface   The maximum
          of $T_c$ is at a field
          $H_R \sim 1.5 H_L$. 
                                   }
\label{Fig1}
\end{figure}

Among these systems,  URhGe has attracted much attention  both experimentally \cite{LevyScience05,Hardy05,Miyake08,HuxleyLifshitz11,HardyPRB11, Aoki11,Aoki14, Aoki142,Fujimori14,Tokunaga15,Gourgout16,Wilhelm17,Braithwaite18} and theoretically \cite{Hattori13, Mineev15,Mineev17}.
  In zero applied magnetic field, it displays ferromagnetism with  magnetic moment  oriented  along the  $c$ -axis, and
   spin-triplet superconductivity at a lower temperature \cite{Hardy05}.
     In an external magnetic field along the $b$-axis ($\mathbf b\perp \mathbf c$), superconductivity disappears at about $B$=2T.
      This is believed to be caused by the orbital effect of the field~\cite{Hardy05}.
     However, at higher magnetic fields, in the range from 8T to 13.5T, it reappears again \cite{LevyScience05} (see  Fig.\ref{Fig1}).

Ferromagnetic spin fluctuations are believed to provide the pairing glue for superconductivity in a ferromagnetic metal
\cite{Mineev15,Mineev17}.
 Indeed,  NMR spin-spin relaxation measurements indicate
 that uniform longitudinal spin fluctuations (the ones in the direction of a magnetic field) are strongly enhanced in the
  field range where superconductivity has been observed \cite{Tokunaga15}.  Measurements of the specific heat \cite{HardyPRB11,Aoki11}, electric conductivity \cite{Miyake08,HardyPRB11,Tokunaga15,Gourgout16}, and magnetisation \cite{HardyPRB11}
   indicate that the increase of spin fluctuations is accompanied by the increase of the effective mass of fermions.  This is indicative of a critical behavior near a ferromagnetic instability.

  In this communication we address the origin of the ferromagnetic instability in a finite field.
   We argue that it is due to a Lifshitz transition observed~ \cite{HuxleyLifshitz11}  in one of the bands which form the
   electronic structure of URhGe.  This Lifshitz transition pushes the system closer to the magnetic instability and  enhances the magnetic fluctuations.
    This in turn  leads to re-entrance of superconductivity (RSC) at a finite field.

 The minimal model of the electronic structure of  URhGe has 
 two bands with 
  non-equal dispersions and band minima shifted by  ${\bf K}_0$.
  (see Fig. \ref{Fig1}). 
In a ferromagnetic state at zero field  spin-up and spin-down states in both bands are split by an effective exchange field.
 Both branches of band 1 cross the chemical potential $\mu$, while both branches of band  2 are above $\mu$ (see the left inset in Fig. \ref{Fig1}).
At a finite $H$  the bands experience additional
Zeeman splitting. The exchange field was reported~\cite{LevyScience05} to be rather weak
($\sim 0.1 T$), hence
at fields near 10T  Zeeman splitting dominates.  The dispersions of the two bands with
 Zeeman splitting are $\epsilon_{1,\sigma}(\mathbf k)=\frac{\mathbf k^2}{2 m_1}-\mu-\sigma \mu_BH$ and
$\epsilon_{2,\sigma}(\mathbf k)=\frac{(\mathbf k-\mathbf K_0)^2}{2 m_2}-\mu+\mu_0-\sigma \mu_BH$.
 As the  field increases, the splitting grows, and at some critical field the system experiences a Lifshitz transition, in which
 spin-up branch of band 2 crosses  the chemical potential (the middle inset in Fig. \ref{Fig1}).
We add Hubbard four-fermion repulsive interaction $U$ and 
analyze the tendency towards magnetic order and magnetically-mediate superconductivity in this field range. The parameters relevant to URhGe are presented in~\cite{parameters} 
 \begin{figure}[h]
\includegraphics[width=0.48\textwidth]{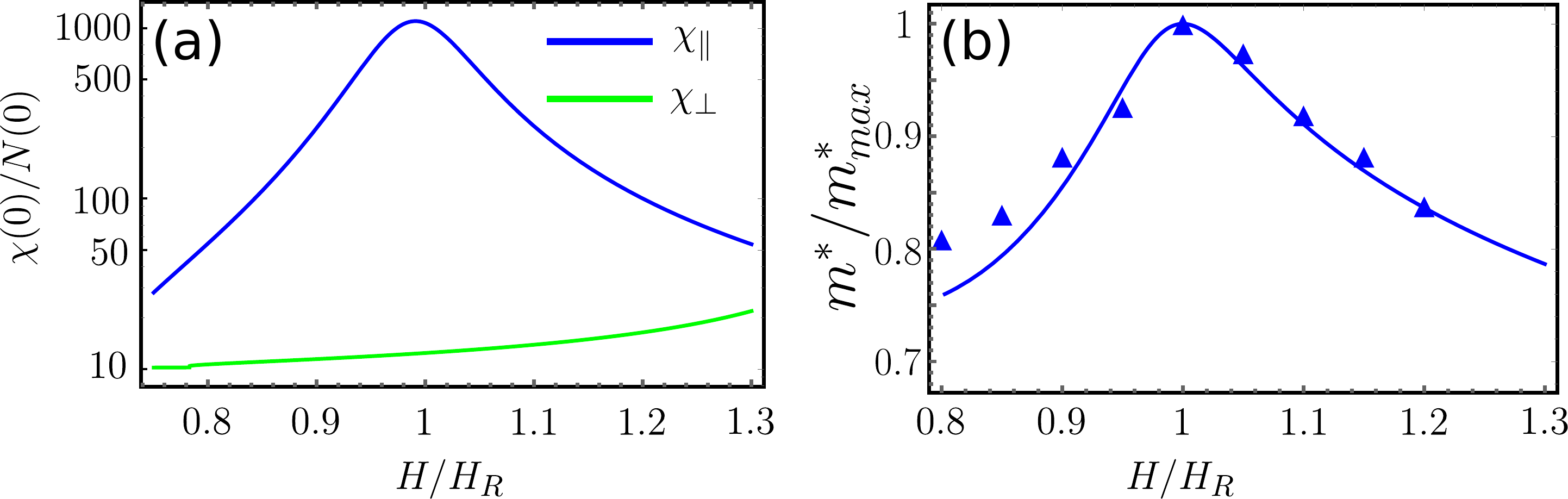}
\caption{(a) Longitudinal (blue/dark grey) and transverse (green/light grey) spin susceptibilities as functions of external magnetic field $H/H_R$. (b) The theoretical results  for the effective mass of fermions in sub-band $1\downarrow$, $m^*_{1\downarrow}(H)$  (normalized to its maximum value at $H \approx H_R$) (solid line) along with $m^*$ 
  extracted from the measurements of magnetisation (triangles).
   \cite{HardyPRB11}. 
   }
\label{Fig2}\end{figure}

We first compute the longitudinal  and transverse  susceptibilities at $T=0$.  Within
 RPA we have
  \cite{Brinkman68}:
  \begin{eqnarray}
  \chi_ {\parallel}(q)&=&\sum_{\sigma}\frac{\chi_0^{\sigma}(q)}{1-U^2\chi_0^{\sigma}(q)\chi_0^{-\sigma}(q)}\label{parrChi},\\
  \chi_ {\perp}(q)&=&\sum_{\sigma}\frac{\chi_0^{\sigma,-\sigma}(q)}{1-U\chi_0^{\sigma,-\sigma}(q)}\label{transChi}.
  \end{eqnarray}
 where  $\chi_0^{\sigma}=\sum_{i=1,2}\chi_0^{i,i,\sigma,\sigma}$ and $\chi_0^{\sigma,-\sigma}=\sum_{i=1,2}\chi_0^{i,i,\sigma,-\sigma}$ are particle-hole susceptibilities of free fermions from the two bands:
 \begin{equation}
 \chi_0^{j,i,\sigma,\sigma'}(\mathbf q,i\omega)= \int_lG_{j,\sigma}(\mathbf l,i\xi)G_{i,\sigma '}(\mathbf l+\mathbf q,i(\omega+\xi))\label{chi},
  \end{equation}
 where $G_{j,\sigma}(\mathbf p,i\omega)=(i\omega-\epsilon_{j\sigma}(\mathbf p))^{-1}$ and  $\int_l...=\int d\xi\frac{d^3 l}{(2\pi)^4}...$.

The results are shown in Fig. \ref{Fig2} a. We see  that the uniform longitudinal susceptibility is  enhanced in the vicinity of the field $H_R$, which is somewhat larger than
 $H_L$, at which the Lifshitz transition occurs ($H_R \sim 1.5 H_L$).   The non-monotonic behavior of $\chi_\parallel (q=0)$ can be understood by noticing that the denominator in (\ref{parrChi}) behaves as
 \begin{equation}
 D=1-U^2(N_{2\uparrow}(0)+N_{1\uparrow}(0))N_{1\downarrow}(0),\label{D}
 \end{equation}
 where $N_{i\sigma}(0)$ is the density of states at the Fermi surface of a sub-band $i =1,2$ with spin projection $\sigma$.  At $H > H_L$, $N_{2\uparrow}(0)$ becomes non-zero, $D$ decreases and $\chi_\parallel (q=0)$ increases. At higher fields $N_{1\downarrow}(0)$ decreases (see right insert in Fig. \ref{Fig1}) and  $\chi_\parallel (q=0)$ decreases.
  At even higher magnetic field the spin-down sub-band of band 1 undergoes the second Lifshitz transition and becomes  unoccupied, in agreement with \cite{HuxleyLifshitz11}.
 The transverse $ \chi_ {\perp}(0)$ has much weaker dependence on $N_{2\uparrow}(0)$ and does not show a peak around $H_R$.  This agrees with
  the NMR results \cite{Tokunaga15}.

 \begin{figure}[h]
\includegraphics[width=0.49\textwidth]{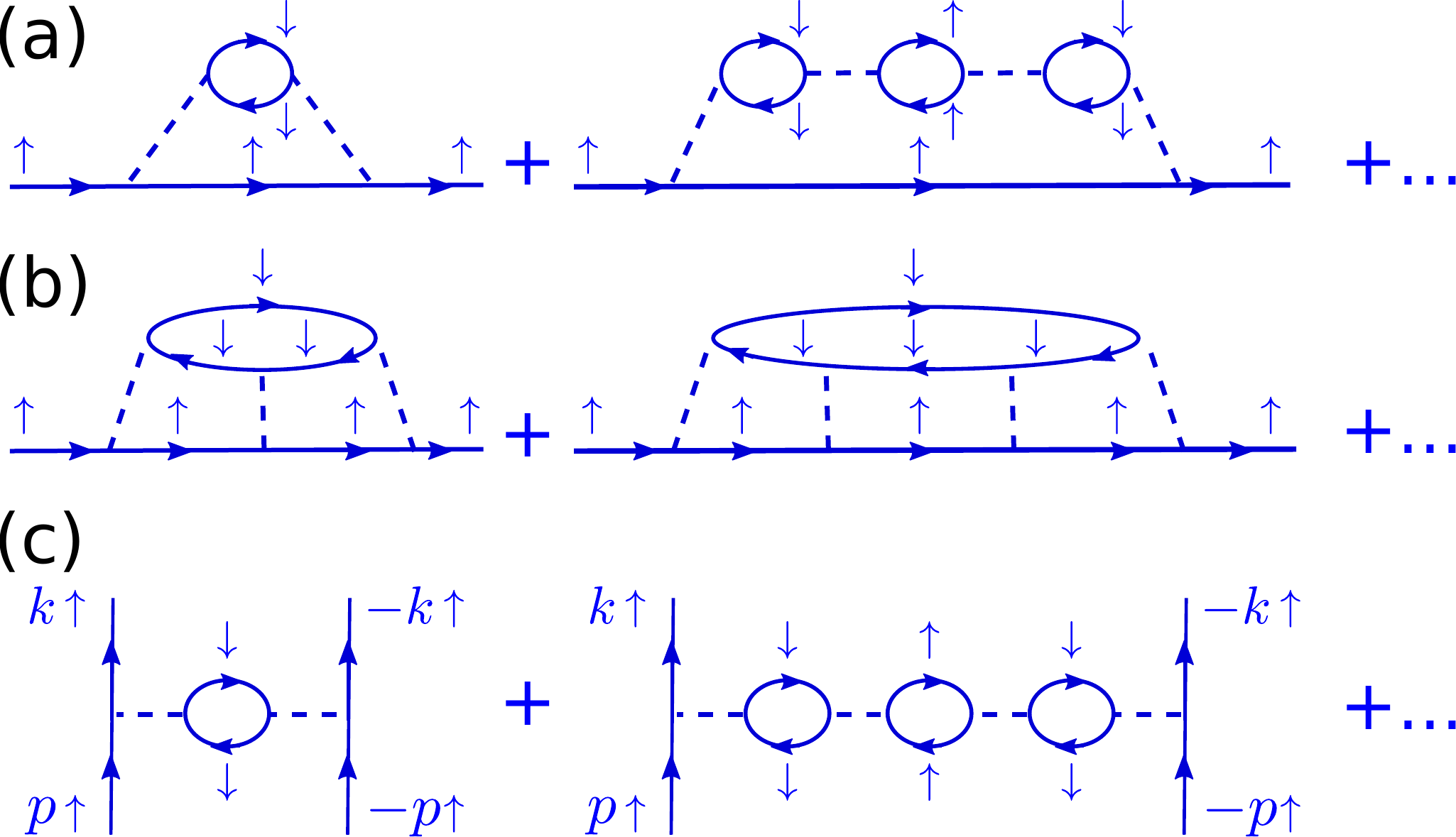}
\caption{Feynman diagrams describing the first two orders of the RPA series  for (a) longitudinal component of the self-energy, $\Sigma_{i\uparrow}^L$; (b) transverse component of the self-energy, $\Sigma_{i\uparrow}^T$; and (c) p-wave pairing vertex. }
\label{Fig3}\end{figure}

 The effective mass of a conduction electron is given by
 \begin{eqnarray}
 \frac{m^*_{i\sigma}}{m_{i\sigma}}=[1-\partial_\omega\Sigma_{i\sigma}(\mathbf p_{i\sigma F},\omega)]_{\omega=0},
 \end{eqnarray}
 The electron self-energy $\Sigma$ can be written as a sum of longitudinal and transverse components.  The corresponding ring and ladder diagrams \cite{Brinkman68,Fay80} are shown in Fig. \ref{Fig3} a and b, respectively:
 \begin{eqnarray}
 \Sigma_{i\sigma}^L(\mathbf p,i\omega)&=&\int_qG_{i\sigma}(\mathbf p-\mathbf q,i(\omega-\xi))V_L^{-\sigma}(\mathbf q,i\xi),\label{SigL}\\
 \Sigma_{i\sigma}^T(\mathbf p,i\omega)&=&\int _qG_{i,-\sigma}(\mathbf p-\mathbf q,i(\omega-\xi))V_T^{\sigma,-\sigma}(\mathbf q,i\xi),\label{SigT}
 \end{eqnarray}
 where the effective interactions are
 \begin{eqnarray}
 V_L^{-\sigma}(\mathbf q,i\omega)&=&\frac{U^2\chi_0^{-\sigma}(\mathbf q,i\omega)}{1-U^2\chi_0^{-\sigma}(\mathbf q,i\omega)\chi_0^{\sigma}(\mathbf q,i\omega)},\label{VL}\\
 V_T^{\sigma,-\sigma}(\mathbf q,i\omega)&=&\frac{U^3[\chi_0^{\sigma,-\sigma}(\mathbf q,i\omega)]^2}{1-U\chi_0^{\sigma,-\sigma}(\mathbf q,i\omega)},\label{VT}
 \end{eqnarray}
 Because the susceptibility $\chi (q)$ is enhanced at $q=0$, the largest contribution to the self-energy
  comes from intra-band scattering  (i.e., the  scatterng within band 1 or band 2), while inter-band interactions contribute much less
  (see Supplemental Material (SM)). Performing frequency integration and following \cite{Brinkman68,Fay80} we obtain
 \begin{eqnarray}
 \frac{m_{i \sigma}^*}{m_{i \sigma}}=1+\lambda_L^{i\sigma}+\lambda_T^{i\sigma},\label{mass}
 \end{eqnarray}
 where
 \begin{eqnarray}
\lambda_L^{i\sigma}&=&\frac{m_i}{(2\pi)^2p_{Fi\sigma}}\int_0^{2p_{Fi\sigma}}qdq V_L^{-\sigma}(\mathbf q,0),\label{lon}\\
\lambda_T^{i\sigma}&=&\frac{m_i}{(2\pi)^2p_{Fi\sigma}}\int_{p_l}^{p_u}qdq V_T^{\sigma,-\sigma}(\mathbf q,0).\label{trans}
   \end{eqnarray}
  Here $p_{Fi\sigma}$ is the Fermi momentum of the sub-band $\{i,\sigma\}$, and the integration limits for the transverse component are  $p_l=max\{p_{Fi\sigma}-p_{Fi,-\sigma},0\}$ and $p_u=p_{Fi\sigma}+p_{Fi,-\sigma}$.

\begin{figure}[h]
\includegraphics[width=0.48\textwidth]{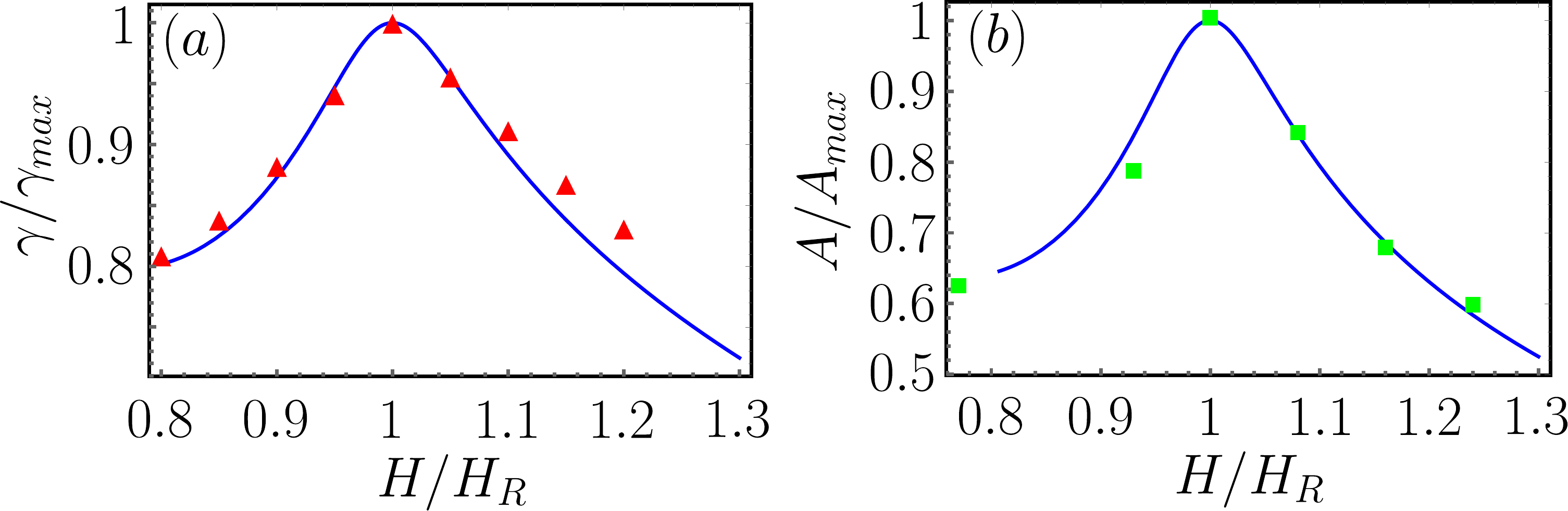}
\caption{Enhancement of (a) Sommerfeld coefficient $\gamma$  measured in \cite{HardyPRB11} (triangles) and (b) conductivity coefficient $A$  measured in Ref \cite{Gourgout16} (squares) and calculated with our model (solid line) for the same parameters as in Fig. \ref{Fig1}. }
\label{Fig4}\end{figure}

The result of the calculation of $m_{i \sigma}^*/m_{i \sigma}$ for  the sub-band $1\downarrow$ is shown in Fig.  \ref{Fig2} b. As expected, the  mass enhancement is
 peaked at $H \approx H_R$, where  the uniform susceptibility is the largest.   The effective masses for other sub-bands show similar enhancement (see SM).
The theoretical result $m^*/m$  agrees well with the mass ratio extracted from magnetisation measurements \cite{HardyPRB11} (see  Fig. \ref{Fig2} b).
Using the result for $m^*(H)$, we computed the specific heat and resistivity.  The main contribution to the specific heat comes from the $1\downarrow$ sub-band, and the Sommerfeld coefficient $\gamma = C(T)/T$  can be estimated as~\cite{Jacko09} $\gamma\propto N_{1\downarrow}(0)m_{1\downarrow}^{*}/m_{1}$. In Fig. \ref{Fig4}a we show the calculated $\gamma$ and the experimental one from Ref.~\cite{HardyPRB11}. Clearly, both are peaked around $H_R$ and show similar behavior at smaller and higher fields.  In Fig. \ref{Fig4} b we show theoretical and experimental results for the prefactor $A$ in the expression for the resistivity $\rho = A T^2$. Theoretical $A$  has been obtained using the
 Kadowaki - Woods relation \cite{Kadowaki86} $A/\gamma^2=const$, the experimental results are from Ref. ~\cite{Gourgout16}.  Again, the agreement is quite good.

We next turn to superconductivity.
The reduction and subsequent disappearance of superconductivity at small fields has been argued to be the orbital effect of a field~\cite{Scharnberg80,Hardy05}.
The reduction of $T_c$ by a vector potential follows~\cite{Scharnberg80}
\beq
T_c (H) = T_c (H=0)  \left(1 - H/H_c\right),  ~~~ H_c = \frac{20\pi k_B^2 T_c^2(0)}{7\zeta (3)e\hbar v_F^2\mu_0}
\label{new_1}
\eeq
 This form of $T_c (H)$ agrees with the data at small fields~\cite{Hardy05}.  When $H > H_c$, $T_c (H)$ vanishes.

To study the re-entrant superconductivity, we do the analysis in two steps. First, we compute $T^{Eli}_c$  within Eliashberg spin-fluctuation formalism,
without including the orbital effect of a field.   We then use the result for $T^{Eli}_c$ as an effective $T_c (0)$ to estimate $H_c$ from (\ref{new_1}). We then use  (\ref{new_1}) to obtain the actual $T_c$.

An exchange by ferromagnetic spin fluctuations enhances the pairing vertex in p-wave channel, and
 below we search for superconducting order with p-wave symmetry.  We analyze all fields and keep both Zeeman and exchange splitting.
   To keep calculations under control, we neglect the feedback from ferromagnetic order on the pairing interaction.
   This feedback is relevant near a ferromagnetic quantum-critical point~\cite{Haslinger}, but less relevant  away from criticality, where our analysis holds.

In Eliashberg theory one needs to solve the set of equations for quasiparticle $Z_{i\sigma}(\mathbf q,i\omega_n) = 1 - \Sigma_{i\sigma}(\mathbf q,i\omega_n)/(i\omega_n)$ and the
 pairing vertex, $W_{i\sigma}$:
\begin{equation}
(1-Z_{i\sigma}(p))i\omega_n=T\sum_{p'}\frac{V_z^{-\sigma}(p-p')Z_{i\sigma}(p')i\omega_{n'}}{(i\omega_{n'}Z_{i\sigma}(p'))^2-\epsilon_{i\sigma}(\mathbf p')^2},\label{El1}
\end{equation}
\begin{equation}
W_{i\sigma}(p)=T\sum_{p'}\frac{V_{W}^{-\sigma}(p-p')W_{i\sigma}(p')}{(i\omega_{n'}Z_{i\sigma}(p'))^2-\epsilon_{i\sigma}(\mathbf p')^2},\label{El2}
\end{equation}
 where $\sum_{p'}...=\sum_{\omega_{n'}}\int d^3p'...$,  $\omega_n=\pi(2n+1)T$, $p=\{\mathbf p,i\omega_n\}$, and the interactions are $V_z^{-\sigma}=V_L^{-\sigma}+V_T^{\sigma,-\sigma}$ and $V_{W}^{-\sigma}=V_L^{-\sigma}$.
We use a standard trick and reduce Eliashberg set to a single equation by introducing  $\Phi(p)=W(p)/|\omega_nZ(p)|$.  Then $\Phi(p)$ is expanded in spherical harmonics and only the
p-wave piece, $\Phi_{i\sigma}^1$ is retained.
Integrating over momenta as $\int d^3p...=\int d\Omega\int d\epsilon N_{i\sigma}(0)...$, where $\Omega$ is the solid angle, we obtain an integral equation of $\Phi_{i\sigma}^1 (\omega_n)$ in the form
\begin{equation}
\sum_{n\geq 0}K_{mn} (\omega_m \omega_n) \Phi_{i\sigma}^1(\omega_n)=0,\label{Kmn}
\end{equation}
where
\begin{eqnarray}
&&K_{mn}=\lambda_{-\sigma}^{(1)}(\omega_m-\omega_n)+\lambda_{-\sigma}^{(1)}(\omega_m+\omega_n)\nonumber\\
&&-\delta_{mn}\left |\sum_{l=-N-1}^{N}\lambda_{-\sigma}^{(0)}(\omega_m-\omega_l)sgn(\omega_l)+\frac{\omega_m}{\pi T}\right |,\label{K}
\end{eqnarray}
We introduced $\lambda_{-\sigma}^{(1)}=-N_{i\sigma}{(0)}\int d\Omega Y_1(\cos(\theta))V_{W}^{-\sigma}(\theta)$ and  $\lambda_{-\sigma}^{(0)}=N_{i\sigma}{(0)}\int d\Omega V_{z}^{-\sigma}(\theta)$, where $Y_1 (\theta)$  is the first spherical harmonic and $\theta$ is the angle between $\mathbf p$ and $\mathbf p'$.

\begin{figure}[h]
\includegraphics[width=0.48\textwidth]{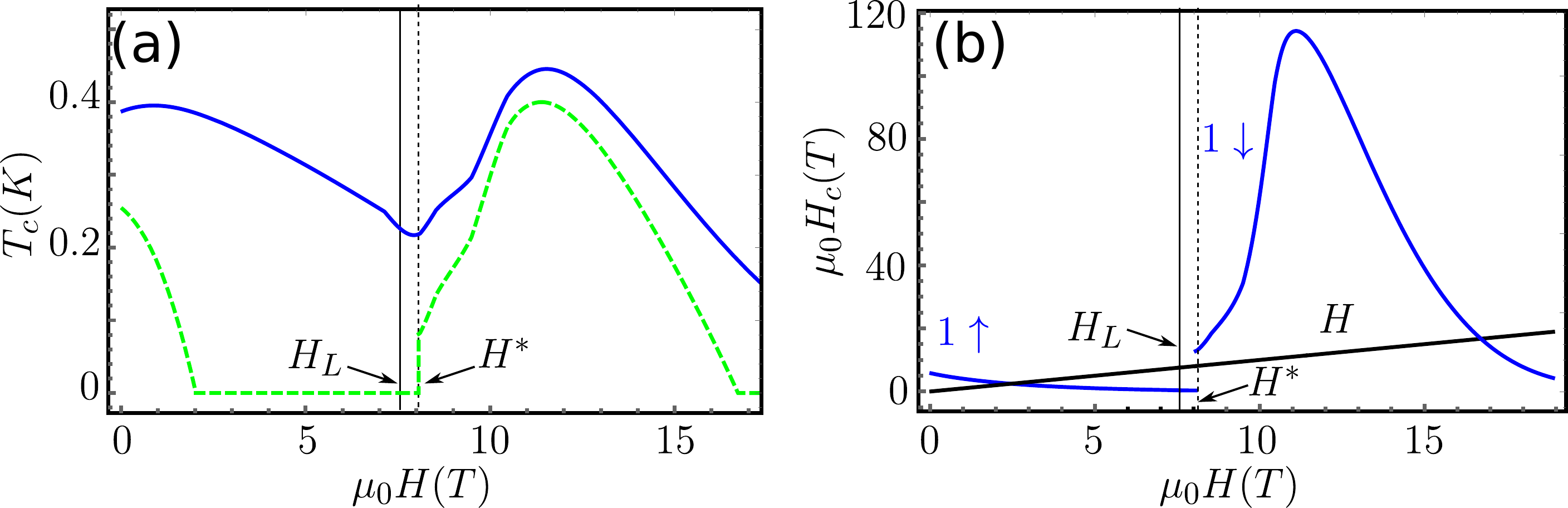}
\caption{(a) Solid line -- $T^{Eli}_c (H)$, obtained within Eliashberg formalism without including the orbital effect of the field.
Dashed line -- the actual $T_c$, with both Zeeman/exchange and orbital effects. The actual $T_c$ is always smaller than $T^{Eli}_c (H)$  due to the orbital effect of the external and the exchange fields.
 (b) The effective field $H_c$ from Eq. \ref{new_1}  as a function of $H$.  Orbital effect destroys superconductivity when $H_c$ (blue line) is smaller than $H$ (thin black line). The value of $H_c$ changes discontinuously at $H= H^*$, when superconductivity switches from  sub-band $1\uparrow$ to sub-band $1\downarrow$, where the effective mass is larger. This gives rise to a jump in the actual $T_c$ in panel (a).
 }
\label{Fig5}\end{figure}

Keeping only the interaction with small momentum transfer, we factorize the pairing between three bands:  up and down sub-bands of band 1 and
spin-up sub-band of band 2. We recall, however, that effective p-wave pairing interaction between fermions on a given band is the sum of contributions from particle-hole bubbles from all three bands.   We solve   Eq. (\ref{K}) for all three bands and find the largest $T_c$ (see SM for details). The result
is shown as a solid line in  Fig. \ref{Fig5}a.  At smaller fields  $H < H^*$, where $H^*$ is slightly above $H_L$,
 superconductivity develops on the $1\uparrow$ sub-band.  At  at $H > H^*$  it switches to sub-band $1\downarrow$, and $T_c$ for superconductivity on this band has a maximum at $H \sim H_R$, where the effective mass on this band is also maximal.
We next include the orbital effect. In Fig.  \ref{Fig5}b we show $H_c$ from Eq. (\ref{new_1}) as a function of external $H$. Orbital effect destroys  superconductivity when $H_c <H$.  We see that this holds at intermediate fields, in the range where without orbital effect superconductivity would develop at  sub-band $1\uparrow$.  At higher fields, superconductivity switches to  sub-band $1\downarrow$, where the effective mass and $T^{Eli}_c (H)$  are larger, and $k_F$ is smaller. Each  modification increases $H_c$, which becomes larger than $H$, in which case orbital effects do not destroy superconductivity. We show actual $T_c$ by dashed line in Fig. \ref{Fig5} b and by solid line in Fig. \ref{Fig1}. Note that $T_c$ appears discontinuously at $H = H^*$, where $T^{Eli}$ switches from $1\uparrow$ to  $1\downarrow$ sub-band. Inter-band pairing interactions likely
 smoothen the first-order phase transition.  The theoretical profile of $T_c$ vs $H$ agrees nicely with the data~\cite{LevyScience05} (see Fig. 1).

To summarize, in this communication we argued that the enhancement of the effective mass in URhGe at fields near 10 T and the emergence of RSC around this field are due to
 Lifshitz transition. We considered the model for URhGe with two electronic bands and analyzed the behavior of the system near a field when the bottom of the  spin-up branch of previously unoccupied band 2  sinks below the Fermi level.
 We first computed  transverse and longitudinal spin susceptibilities and argued that the longitudinal susceptibility dramatically enhances in some field range above the Lifshitz transition, while the transverse susceptibility remains flat.
 This fully agrees with the  behavior of longitudinal and transverse susceptibilities, extracted from NMR measurements of the relaxation times, $1/T_1$ and $1/T_2$ \cite{Tokunaga15}.
We next computed the one-loop self-energy due to magnetically-mediated interaction and obtained the  enhancement of the effective mass. The theoretical result for $m^*/m$ agrees with the experimental data extracted from  magnetisation measurements~\cite{HardyPRB11,Note111}. We found good agreement also for
  the Sommerfeld coefficient  and the prefactor for the $T^2$ term in the resistivity~ \cite{HardyPRB11,Gourgout16}.
   We next analyzed superconductivity. We first solved the Eliashberg equation for magnetically-mediated superconductivity without orbital effect of a field and obtained $T^{Eli}_c$ with a  maximum at a field where the effective mass is the largest.  Superconductivity resides on $1\uparrow$ sub-band at smaller fields and on  $1\downarrow$ sub-band at higher fields. We then added addition  pair-breaking orbital effect and found that superconductivity
   exists at small fields, gets destroyed by orbital effect at intermediate fields, and re-appears discontinuously roughly at a field of Lifshitz transition.
   This behavior fully agrees with the data~\cite{LevyScience05} (Fig. \ref{Fig1}). The reduction of theoretical $T_c$ at higher fields is somewhat slower than in the data.  The reason could be
a re-orientational  transition, detected at 12T  \cite{LevyScience05}, in which the magnetic moment  rotates towards the field direction,
    leaving its  magnitude  unchanged. This spin re-orientation does not  increase  longitudinal fluctuations but complicates the field dependence of $T_c$ above 12T. Overall, it looks increasingly likely that  topological Fermi-surface transitions can account for much of the puzzling physics in nearly magnetic itinerant systems \cite{SlizovskiyChubukovBetouras, SlizovskiyBetourasCarrQuintanilla}.
\begin{acknowledgments}
We thank R. Fernandes and M. Greven  for  useful conversations.
This work was supported by the EPSRC (YS and JJB) through the grant EP/P002811/1.
and by the U.S. Department of Energy through
the University of Minnesota Center for Quantum Materials,
under award DE-SC-0016371 (A.V.C.).
\end{acknowledgments}
\bibliography{ybs.bib}{}

\begin{thebibliography}{26}
\bibitem[{par()}]{parameters}
\bibinfo{note}{The parameters of the electronic structure are $m_1=8.06\times
  10^{-28}kg$, $m_2=2.4 m_1$, $v_{F0}=10434m/s$, $\mu_0/\epsilon_F^0=$1.48. For
  calculation of superconductivity we used $U N(0)=1.03$. Here
  $\epsilon_F^0=[3n(2\pi)^2/4]^{2/3}/(2m_1)$,
  $v_{F0}=\sqrt{\frac{2\epsilon_F^0}{m_1}}$, and
  $N(0)=\frac{m_1^{3/2}(\epsilon_F^0)^{1/2}}{2^{1/2}\pi^2}$ are the Fermi
  energy, Fermi velocity and density of states of band 1 in the absence of exchange and
  external magnetic fields. $n$ is the total electron density}.
\end{thebibliography}
\bibliographystyle{apsrev}

\end{document}